# Interferometric scattering enables fluorescence-free electrokinetic trapping of single nanoparticles in free solution

*Allison H. Squires[1], Abhijit A. Lavania[1,2], Peter D. Dahlberg[1], and W.E. Moerner[1,2,]\**

[1] Department of Chemistry, Stanford University, Stanford, California 94305

[2] Department of Applied Physics, Stanford University, Stanford, California 94305

ABSTRACT: Anti-Brownian traps confine single particles in free solution by closed-loop feedback forces that directly counteract Brownian motion. The extended-duration measurement of trapped objects allows detailed characterization of photophysical and transport properties, as well as observation of infrequent or rare dynamics. However, this approach has been generally limited to particles that can be tracked by fluorescent emission. Here we present the Interferometric Scattering Anti-Brownian ELectrokinetic trap (ISABEL trap), which uses interferometric scattering rather than fluorescence to monitor particle position. By decoupling the ability to track (and therefore trap) a particle from collection of its spectroscopic data, the ISABEL trap enables confinement and extended study of single particles that do not fluoresce, that only weakly fluoresce, or which exhibit intermittent fluorescence or photobleaching. This new technique significantly expands the range of nanoscale objects that may be investigated at the single-particle level in free solution.

TEXT: The consequence of Brownian motion in solution-phase single-molecule fluorescence spectroscopy experiments is that nanoscale objects quickly diffuse out of focus or through a confocal observation volume, resulting in brief bursts of signal that can reveal population-wide heterogeneity or sub-millisecond dynamics.[1-5] In order to access longer timescales or record infrequent events, it is necessary to employ methods that prolong observation of individual particles by overcoming diffusion. One such strategy is to actively maintain the position of the particle within an observation volume in free solution using closed-loop feedback, either by quickly following the trajectory of the particle with a positioning stage,[6-10] or by applying forces that counteract the effects of Brownian motion.[11] The latter class of approaches, collectively known as anti-Brownian traps,[12-14] can circumvent the potential risks of perturbation due to interactions with or alterations of the local nano-environment that may accompany immobilization of particles by tethers, surface attachment, or encapsulation.[15-21]

All implementations of anti-Brownian traps can be distilled to two essential aspects of the closed-loop feedback: first, real-time tracking provides the location of a particle relative to the target position, and may be determined using fluorescence[6,11] or bright- or dark-field imaging[8,22,23] in combination with either camera-based tracking[23-25] or timed movement of the excitation beam and one or more point detectors.[6,9,26-28] Second, a feedback force must be quickly applied to move the particle back towards the target, which may be implemented using electric fields to induce electrophoresis or electroosmosis,[28-30] thermal gradients to induce thermophoresis,[31] optical forces,[32] or differential pressure to induce hydrodynamic flow.[33] These steps must be implemented quickly enough to overcome the diffusive motion of the particle, and significant recent progress has been made toward optimizing this control loop to enable trapping of individual small organic fluorophores.[28,34] Trap implementations that utilize either intrinsic or label-based fluorescence to track emissive particles have been employed to characterize time-varying photophysical states,[35-40] molecular dynamics and kinetics,[41-47] and more.[48-50] However, the trapping duration, and therefore data collection, in anti-Brownian traps is typically limited by photobleaching or by blinking because dark particles cannot be tracked and are quickly lost. Bechhoefer and co-workers successfully demonstrated trapping of non-fluorescent particles using a dark-field signal,[23] but the unfavorable scaling of scattering intensity with particle size limits this approach to relatively large nanoscale objects (>100 nm) with large scattering cross-sections.

We introduce here the Interferometric Scattering Anti-Brownian ELectrokinetic (ISABEL) Trap, a new anti-Brownian device to track and trap small, non-fluorescent nanoscale objects by using the interference signal between scattered light from the particle and a constant reference field to rapidly estimate a nanoparticle's position. Closed-loop electrokinetic feedback can then be used to control the position of the particle. We demonstrate trapping of gold, polystyrene, and semiconductor nanoparticles as small as 15-20 nm in diameter, and show that the ISABEL trap completely decouples the ability to trap a nanoparticle from measurements related to its fluorescence. Because the interferometric scattering signal scales favorably with particle size relative to scattering alone, the ISABEL trap significantly broadens the range of trappable objects to include small nanoparticles with weak, highly variable, or even no fluorescent signal.

The ISABEL trap is conceptually similar to its Anti-Brownian Electrokinetic (ABEL) trap predecessors,[12] with the key modification being that the particle is tracked interferometrically. Recently, interferometric scattering microscopy techniques which utilize a coherent reference (or local oscillator) field, $E_r$, to homodyne detect a scattered electric field, $E_s$, have been developed for detection and tracking of single weak scatterers.[51-53] The favorable scaling of an interferometric scattering signal, (commonly abbreviated as "iSCAT"[54]) enables detection and high-speed tracking of nanoscale particles,[55-57] and has been used to directly detect biological molecules including viruses,[54] cell secretions,[58] and microtubules,[59] and to weigh single proteins.[60] The recent progress in this field motivated us to develop interferometric scattering into a useful signal for active-feedback nanoparticle trapping.

In the ISABEL trap, particles of interest are diluted and loaded into a quartz microfluidic cell (Figure 1a and b), across the trapping region of which a low-coherence length laser is scanned in a 32-point grid pattern to create the incident excitation field, $E_i$, which covers a roughly 2x2 $\mu m^2$ square trapping region (Figure 1b). The reflected light from the quartz-water interface forms a reference field, $E_r$, which along with the scattered field, $E_s$, is collected by a high-NA objective lens and detected on a photodiode, as shown in Figures 1a and c. The reflected and scattered fields are both separated from the excitation beam using a linear polarizing beamsplitter in combination with a quarter waveplate. The beam scanning in $x$ and $y$ is produced by a pair of acousto-optic deflectors (AODs) controlled by a field-programmable gate array (FPGA), so that the beam position is precisely known as a function of time. Thus, the signal detected on the photodiode over the course of the beam scan can be directly mapped to the scan grid. In addition to detection of scattered and reflected light, fluorescence signals may be simultaneously acquired in a separate detection channel on an avalanche photodiode.

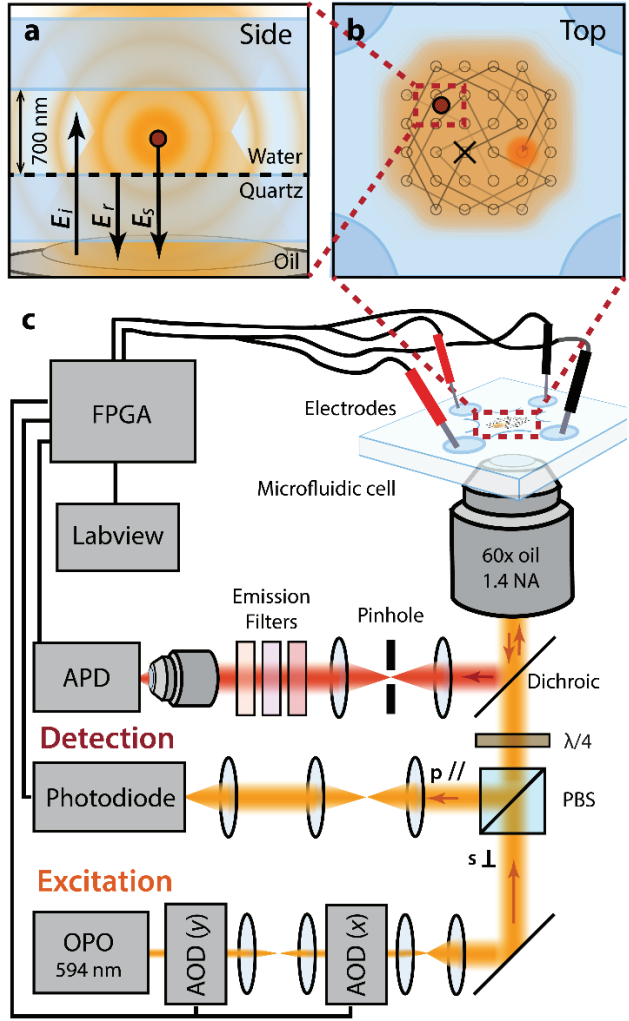

**Figure 1.** The Interferometric Scattering Anti-Brownian ELectrokinetic (ISABEL) Trap. a) A focused incident beam ($E_i$) with low coherence length illuminates a particle in a shallow microfluidic cell ($h = 700$ nm), and the backscattered light ($E_s$) is collected by a high-NA objective. A coherent back-reflection from the quartz-water interface is also collected ($E_r$). b) Top view of microfluidic cell shows the scan pattern, controlled via an FPGA and two acousto-optic deflectors (AODs), of the excitation beam in the center of two crossed microfluidic channels. The trap center, marked by an "X", can be programmatically placed anywhere on this grid. c) Schematic of optical excitation and detection paths for the ISABEL trap. Linearly polarized excitation light passes through a quarter wave plate, so that backscattered and reflected light can be redirected with a polarizing beamsplitter (PBS) and detected at a fast photodiode. Emitted fluorescence can be collected in a separate emission channel. Closed-loop feedback voltages are calculated by the FPGA and applied to the solution using platinum electrodes.

In order to track the particle and thereby determine the appropriate feedback forces necessary to trap it, the particle position is estimated after each complete beam scan and the detected signal is organized into a single 32-point frame (typical frame rates 1-10 kHz). To locate the particle, the photodiode signal at each scan position is first recorded, as shown in Figure 2a. The expected intensity incident upon the detector, $I_{\text{det}}$, depends upon the amplitudes of the reflected and scattered fields $E_{\text{r}}$ and $E_{\text{s}}$, as well as their relative phase, $\theta$:

$$I_{\text{det}} \propto |E_{\text{r}} + E_{\text{s}}|^2 = |E_{\text{r}}|^2 + 2|E_{\text{r}}||E_{\text{s}}| \cos \theta + |E_{\text{s}}|^2 \tag{1}$$

$E_{\text{r}}$ and $E_{\text{s}}$ are generated by interactions of the incident field, $E_{\text{i}}$, with an interface and the scattering particle, respectively. $E_{\text{r}}$ is determined by the reflectivity of the interface, $r$, so that $E_{\text{r}} = r E_{\text{i}}$. Typical values for $r^2$ are on the order of 1%; a glass-air interface reflects 4%, and a glass-water interface reflects 0.4% of normally incident light.[61] $E_{\text{s}}$ is determined by the complex scattering coefficient, $s$, so that $E_{\text{s}} = s E_{\text{i}}$. In the Rayleigh limit, $s$ is proportional to the complex polarizability of the particle, which is described by the particle's volume, $V$, and the complex bulk polarizability of the material relative to the surrounding medium, $\alpha(\lambda)$, where $\lambda$ is the wavelength of the incoming light: $s \propto \alpha(\lambda) \cdot V$.[62]

It is important to note that $E_{\text{i}}$ (which varies spatially in both magnitude and phase due to the focused excitation beam) will generate $E_{\text{s}}$ only at the position of the scattering particle, while the reflected field is generated across the entire beam profile. Considering for the moment an on-axis particle position, where the magnitude of $sE_{\text{i}}$ would be greatest, Equation 1 can be re-written as:

$$I_{\text{det}} \propto |r E_{\text{i}} + s E_{\text{i}}|^2 = |E_{\text{i}}|^2 (r^2 + 2r|s| \cos \theta + |s|^2) \tag{2}$$

Our large area point detector measures the integrated value of the intensity image $I_{det}$, denoted $S_{det}$, for each grid position, and after each complete scan these values are used to reconstruct a single ISABEL image frame, as shown in Figure 2b, where each pixel is assigned the recorded raw signal value from the corresponding scan location (Figure 2b). It is clear from Equation 2 and the definition of $s$ above that the dark-field term, $|E_i|^2|s|^2$, will scale with the square of the particle volume, or with the sixth power of the diameter, $d^6$, and therefore will become negligible in comparison to the homodyne term, $2|E_i|^2 r|s|\cos\theta$, for small particles. The reference term, $|E_i|^2 r^2$, is generated from the quartz-water interface of the microfluidic cell, and usually dominates the measurement. This reflection should remain constant over time, so in order to isolate the desired homodyne term in Equation 2, a background frame containing only the reflection, $S_{bkg}$, is subtracted and used for normalization. For each point in the ISABEL scan, the absolute fractional contrast, $C_f$, is defined as:

$$C_f = \left|\frac{S_{det} - S_{bkg}}{S_{bkg}}\right| \tag{3}$$

The same frames depicted in Figure 2b are shown in Figure 2c as fractional contrast prior to taking an absolute value, and demonstrate that the homodyne term may take either positive or negative values at different points in the scan. The homodyne term can also change sign due to motion of the particle and subsequent change in the relative phase $\theta$. Therefore, to overcome these issues and reliably identify the scan point that deviates most from the background, the absolute value of the fractional contrast is calculated by the FPGA (Eqn. 3 and Figure 2d) and the location of max($C_f$) is used as the particle position in that frame. It is worth noting that the residual Brownian motion of the nanoparticle (even in the axial direction) causes various relative phases to be sampled during each scan position. In spite of this, the absolute fractional contrast in terms of $s$ and $r$ still scales approximately as $\left|\frac{|s|}{r}\right|$.

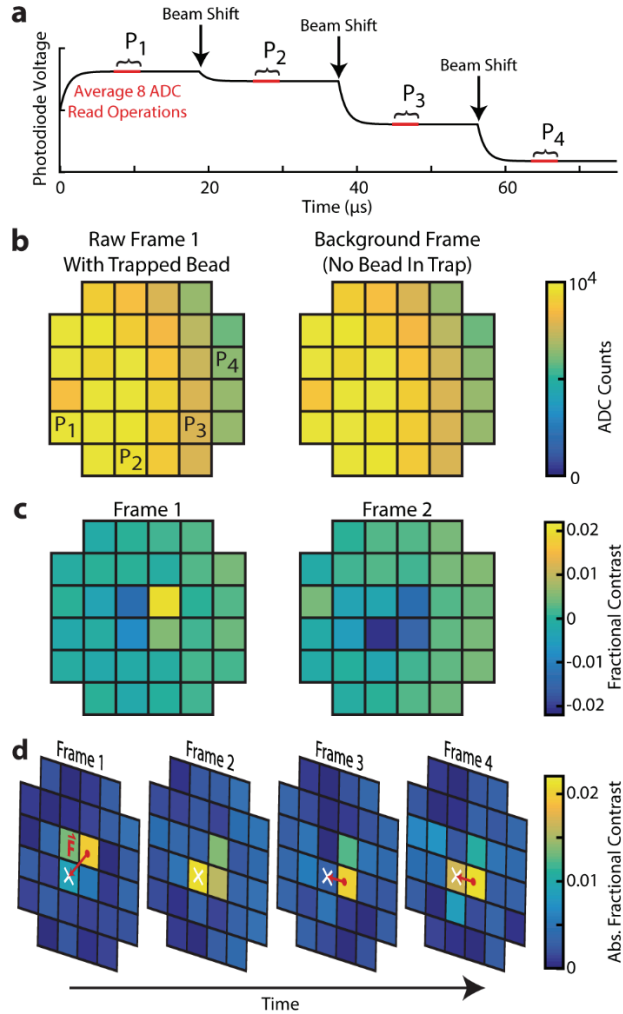

**Figure 2.** Image reconstruction and trapping algorithm. a) The photodiode voltage contains the interference information between the scatterer and the reflected beam, and the signal is digitized at the times $P_i$ shown, after multiple photodiode time constants. b) A representation of the signals recorded from the various scanning beam positions (pixels), also showing the background signal with no bead in trap. c) Flat-fielded fractional contrast signals after removal of background for two complete frames. d) Illustration of the trapping algorithm. After each 600 μs frame, the largest absolute value is used to define the force direction needed to move the particle to the trap center.

After the estimated particle position is determined, the displacement vector to the trap center is calculated, and appropriate feedback voltages are immediately applied to the electrodes in the microfluidic cell by the FPGA for the duration of one frame. As in the previous ABEL trap designs, the resulting applied field is locally uniform with no gradient. The resulting drift force directs the particle toward the pixel marked with an X in Figure 2d, and the amplitude of the applied field is scaled linearly with the distance between the estimated pixel position and the target pixel position. If the estimated position is the same as the target, no voltages are applied. Depending upon surface treatment and zeta potentials, the applied voltages generate either an electrophoretic force (depending upon particle charge) or an electroosmotic flow (no requirement on particle charge) that biases the random diffusion of the scatterer in solution toward the target in the middle of the trapping region.

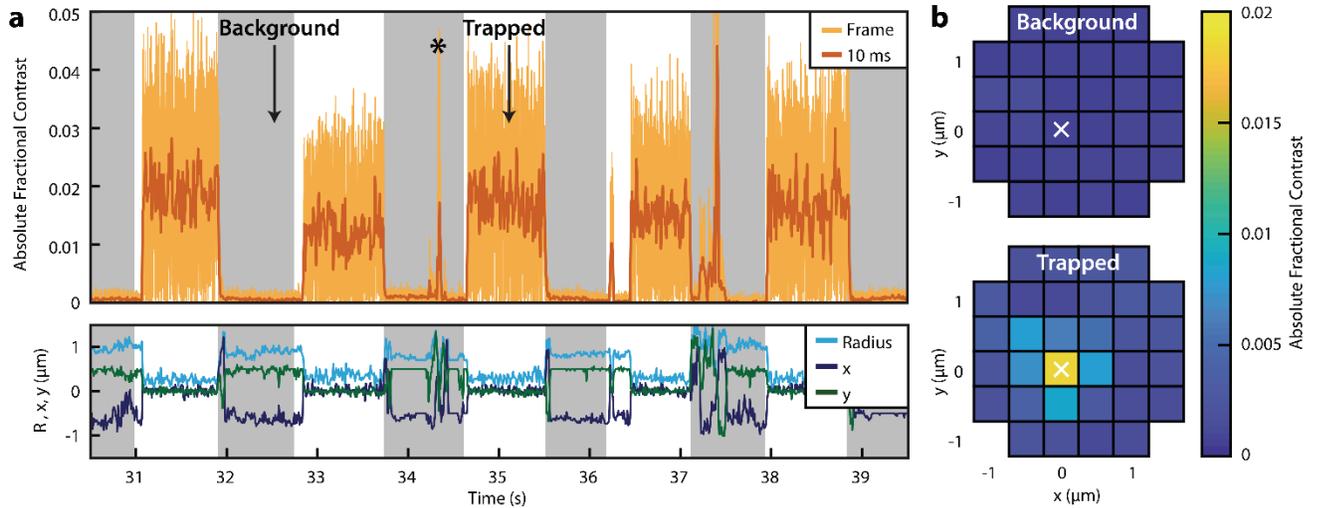

**Figure 3.** Trapping a 40 nm gold nanoparticle. a) Real-time operation of the trap, showing times when the feedback is on (unshaded) and when it is off (gray shading). For a trapping event, the ISABEL signal is shown per frame and also averaged over 10 ms. Asterisk shows a particle diffusing through the trap. The lower curve shows the x, y, and radial position of the maximum $C_f$, averaged in a 10 ms window. b) Empty trap background image averaged over 250 ms (indicated by arrow in a) and image of the trapped object (indicated by arrow in a).

To test the ISABEL trap, we trapped a series of gold nanoparticles of various sizes <100 nm in diameter. Typical results of trapping nominally 40 nm diameter beads can be seen in Figure 3. For each trapped particle, the key variables are the absolute fractional interferometric scattering contrast and the position of the pixel with highest value of contrast defined by $x,y$, and radius $R$ from the software-defined trap center (Figure 3a). In the gray shaded regions, feedback is off and the trap is typically empty, except for occasional diffusion of a bead through the trap, as seen here at ~34.3 s. During these intervals, the values of $x$, $y$, and $R$ are generally random because the algorithm is showing the position of the maximum interferometric scattering signal from noise, which could occur anywhere within the frame. A background frame is typically collected and saved under operator control during this time, for example at the time ~32.5 s in Figure 3a. An x-y plot of the absolute fractional contrast in this background frame is shown in the upper panel of Figure 3b. When feedback is on, a bead is quickly trapped after it diffuses into the trapping region. An example of the spatial distribution of the signal (the "image" of the particle in the trap) for a trapped bead is shown in the lower panel of Fig 3b. It is important to note that for this size bead (~40 nm diameter), the trapping is typically so robust that the feedback must be turned off to release the trapped particle. The real-time plot of fractional contrast in Figure 3a (top) demonstrates that each trapping event exhibits slightly different contrast. These differences are likely due to heterogeneity in bead diameter, which was also observed and quantified by TEM (Figure 4; see also SI Note S1 and SI Figures S3 and S4).

To quantify the functional relationship between bead diameter and absolute fractional contrast, and to assess whether or not the relationship follows the expected linear trend with the cube of the diameter, $d^3$, we separately trapped samples containing gold nanoparticles of five different diameters. Aliquots from the same samples were characterized via transmission electron microscopy (TEM) to quantify particle size (see SI Figures S3 and S4). The mean and standard deviations of the average absolute fractional contrasts for the nanoparticle samples are plotted against the diameters from TEM in Figure 4. The best-fit cubic trend is shown (additional fit details in SI). The scattering contrasts are consistent with the expected $d^3$ scaling for interferometric scattering. By comparison, the expected scaling for the scattering-only signal alone is $d^6$, shown in the dashed line.

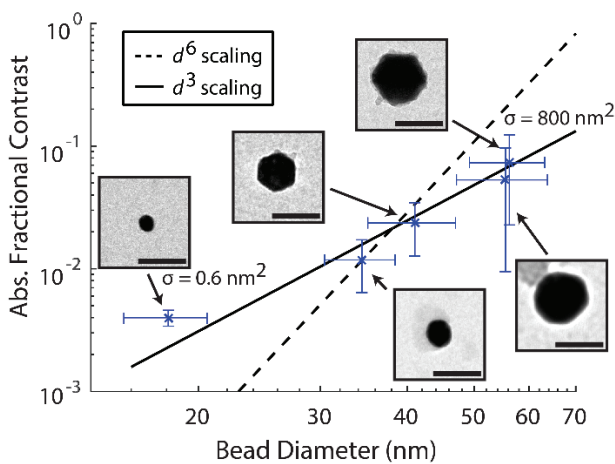

**Figure 4.** The experimentally measured interferometric scattering contrast from trapped gold nanoparticles scales as $d^3$. Gold nanoparticles of nominal diameters 20, 30, 40, 50, and 60 nm were trapped. The diameters of these samples were determined from TEM images (selected images inset, scale bars 50 nm). The symbols and errorbars represent means and standard deviations. The mean calculated scattering cross-sections for the smallest and largest of the gold bead samples are $\sigma = 0.6$ nm$^2$ and $\sigma = 800$ nm$^2$.

To further demonstrate the utility of the ISABEL trap to confine and measure single nanoscale particles that have weak or variable fluorescence, we performed trapping of different types of fluorescent nanoparticles. Figure 5a shows the fluorescent (top) and interferometric scattering absolute fractional contrast traces (bottom) for a trapped fluorescent polystyrene bead (FluoSpheres F8789, ThermoFisher). The expected scattering cross section of this particle is ~1

nm$^2$, comparable to a ~20 nm diameter gold bead, and consistent with the observed contrast of 0.2%. Although the fluorescent signal from the bead photobleaches in just a few seconds under high excitation, the bead remains trapped almost indefinitely – illustrated here by an additional 30 seconds of trapping. In this case, the scattering signal and the fluorescence excitation are produced by the same laser, but it is possible to use two different wavelengths as needed to excite the fluorescence signal in an optimal way. An accompanying video demonstrating continuous trapping of a 50 nm gold particle for several minutes is available as a supplementary video file.

Figure 5b shows the fluorescent (top) and interferometric scattering (bottom) signals for a trapped CdSe semiconductor nanoparticle coated with a thick shell of CdS, with an approximate effective diameter of ~20 nm. In this batch of particles, the observed fluorescence intensity is highly heterogeneous from particle to particle, and exhibits significant emission fluctuations within individual trapping events. Although variability is also observed in the interferometric scattering signal among particles, consistent with the heterogeneous particle morphology (see SI Figure S5), within each trapping event the interferometric scattering signal remains constant.

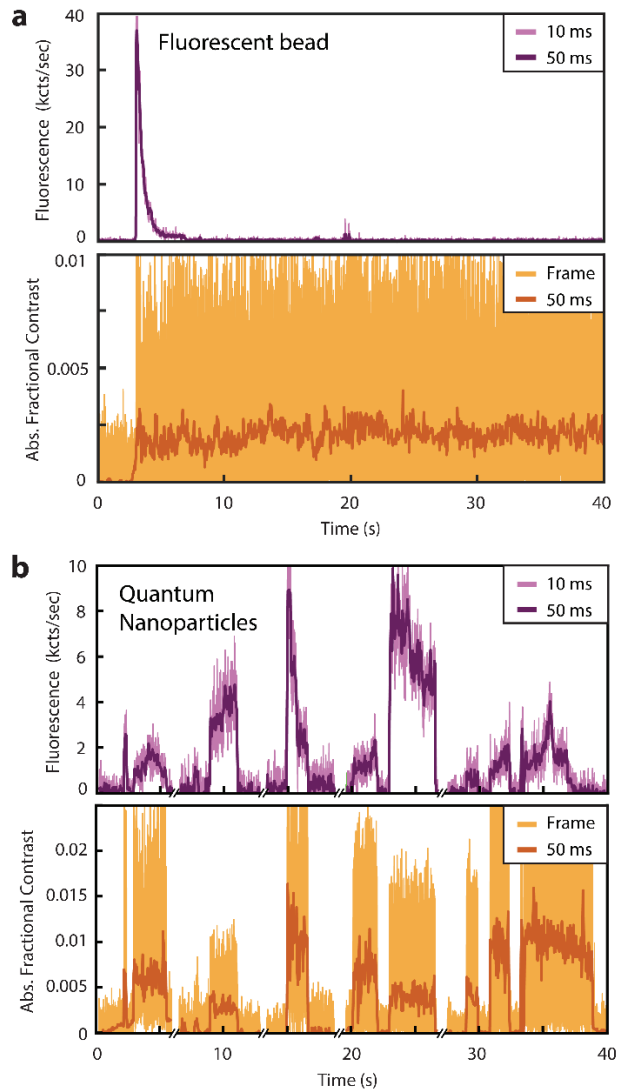

**Figure 5.** The ISABEL trap can confine and measure very weakly fluorescent objects. a) Simultaneous fluorescence and ISABEL signals for a trapped 46 nm fluorescent polystyrene bead. b) Simultaneous fluorescence and ISABEL signals for trapped CdSe/CdS nanoparticles.

In summary, we have demonstrated a single-nanoparticle device, the ISABEL trap, which utilizes interferometric scattering to enable fast position detection and closed-loop feedback trapping of nanoscale particles in solution. In contrast to a dark-field detection approach, the ISABEL trap readily confines objects < 100 nm diameter. With future optimization of trap design and stability, it should be possible to trap single objects as small as those that have been tracked using other interferometric scattering techniques. Critically, the ISABEL trap decouples the ability to trap a particle from spectroscopic or other observations of its nature and photophysical behavior, and therefore permits trapping of a broadly expanded range of nanoscale objects to include those that either do not fluoresce, or that fluoresce only weakly or intermittently. We therefore anticipate that this approach will prove useful for a wide range of future applications in single-molecule biophysical and single-nanoparticle studies in free solution.

ASSOCIATED CONTENT

**Supporting Information**. The following files are available free of charge: Supplementary Information File (.pdf). Includes SI notes, figures, tables, and video captions.

Video S-1: Beam scan pattern (slow) for ISABEL Trap. (.mp4)

Video S-2: ISABEL Trap with 40 nm gold beads (.mp4)

Video S-3: Simultaneous flat-fielded video recording of 40 nm gold beads in the ISABEL Trap. (.mp4)

Video S-4: Minutes-long trapping of a single 50 nm gold bead. (.mp4)

AUTHOR INFORMATION

**Corresponding Author**

* Correspondence and requests for materials should be sent to WEM: wmoerner@stanford.edu.

ACKNOWLEDGMENT

This material is based upon work supported in part by the U. S. Department of Energy, Office of Science, Office of Basic Energy Sciences, Chemical Sciences, Geosciences, & Biosciences Division under Award Number DE-FG02-07ER15892 (Physical Biosciences Program). A.A.L. is an Albion Walter Hewlett Stanford Graduate Fellow. The Jeol JEM1400 TEM was funded by NIH Grant Number 1S10RR02678001 to the Stanford Microscopy Facility.

ABBREVIATIONS

ISABEL trap, Interferometric Scattering Anti-Brownian Electrokinetic trap; ABEL trap, Anti-Brownian Electrokinetic trap; iSCAT, interferometric scattering; FPGA, Field Programmable Gate Array.

**SUPPORTING INFORMATION**

**Interferometric scattering enables fluorescence-free electrokinetic trapping of single nanoparticles in free solution**

Allison H. Squires[1], Abhijit A. Lavania[1,2], Peter D. Dahlberg[1], and W.E. Moerner[1,2,]*

[1] Department of Chemistry, Stanford University, Stanford, California 94305
[2] Department of Applied Physics, Stanford University, Stanford, California 94305

* Correspondence and requests for materials should be sent to WEM: wmoerner@stanford.edu

# Table of Contents

## Supplementary Notes

### Note S1: Materials and sample prep

Samples: Gold beads were purchased from Cytodiagnostics in 20, 30, 40, 50, and 60 nm nominal diameters, and their actual sizes were characterized via TEM (SI Figure S4). Polystyrene beads, 46 nm diameter, were purchased from ThermoFisher (Dark Red FluoSpheres F8789, Ex/Em 660/680 nm). Core-shell CdSe/CdS semiconductor nanoparticles with emission max at 680 nm were fabricated by and purchased from NanoOptical Materials Corporation. The size of these nanoparticles was approximated by their volume as described below, which corresponds to an effective 20 nm diameter.

Metrology: The size distribution of each gold bead sample was determined from TEM micrographs, with representative examples shown in Figure 4. Histograms of the diameters for various bead samples are shown in SI Fig S4. The semiconductor nanoparticles were imaged using cryogenic electron microscopy, with representative micrographs shown in Figure 4. Samples were plunge frozen using a Gatan CP3 plunge freezer on R 2/2 carbon Quantifoil grids and imaged on a Jeol 1400 TEM. The effective area of the semiconductor nanoparticles was calculated from threshold-based binary maps of TEM images, converted to volume, and effective diameters determined corresponding to spherical particles.

Microfluidic cell: The quartz microfluidic cell[1] contains two perpendicular pairs of access channels that intersect at the trapping region, into which the two pairs of electrodes (Pt) for $x$- and $y$-feedback are secured. An outer ring also connects these channels to relieve pressure. The depth of the microfluidic cell in the thin central trapping region was ~700 nm. The magnitude in measured reflection from different microfluidic cells varied two- or threefold (SI Figure S7), likely due to slight variations in fabrication and presence of bonding material acting as a partial surface coating. Cell numbers C9 and Q2 were used most frequently for their consistent performance. Cells were cleaned in piranha solution (3:1 $H_2SO_4$:$H_2O_2$), rinsed thoroughly with DI water, then treated with 1M KOH (10 min) and rinsed again thoroughly prior to use.

Sample preparation: All stock bead samples were sonicated for 10-30 minutes prior to dilution with Nanopure DI water to a working concentration of 1-20 pM, after which they were immediately loaded into microfluidic cells and used for trapping.

### Note S2: Experimental setup

Excitation: A low coherence length laser (Coherent, Mira 900-D mode-locked Ti-sapphire + Mira OPO, 594 nm) is focused to a loose 0.5 μm confocal spot and scanned in a knight's tour pattern[2] across a 32-point, 0.5 μm pitch, ~2x2 μm² trapping region in the center of the microfluidic cell (z-height = 700 nm) to create the ISABEL trap. We found that the short pulses from the laser provided sufficiently low coherence for the experiments. Working powers at the back focal plane range from ~100 μW to ~1 mW. The beam is steered using a pair of AODs (2x AA MT110-B54A1.5-VIS with AA DDSPA2X-D8b15b-34 DDS controller) controlled by an FPGA board (National Instruments PCIe-7842R). The diffraction efficiency changes with AOD voltage, so not all points on the grid are illuminated perfectly evenly, but this was partially mitigated by calibration of the AOD amplitude at each beam position.

Feedback: Two pairs of platinum electrodes are connected to the output of an external 16x amplifier driven by the FPGA, which outputs the appropriate voltage in $x$ and $y$ once the position of the particle is estimated for each frame. Voltages are applied according to a virtual harmonic potential well, such that the applied voltage will increase linearly with the particle's distance from the center of the trap. Typical applied fields range from $10^4$-$10^5$ V/m, and the field produces a constant velocity for the particle for the duration of one frame. If the particle is already at the center, no voltage is applied.

Scattering detection: Scattered and reflected beams are separated from the excitation beam with a polarizing beamsplitter (Thorlabs PBS251, 420-680 nm) and quarter waveplate (WPQ05M-633). Scattered light is detected on a photodiode (90 kHz bandwidth, New Focus Optical Receiver 2031) with active area ~50 mm$^2$. During each experiment, excitation power was optimized so that the detected signal would stay just below saturation of the detector gain setting in use. The photodiode was operated at high gain ($2 \times 10^6$ V/A), and this signal was digitized by the 16-bit ADC of the FPGA.

Fluorescence detection: Fluorescence signals were separated from scattered signals by reflection from a short pass dichroic (Semrock TSP01-704, cutoff set to ~640 nm) and transmission through several optical filters (635 LP, 650-710 BP, 640 LP), as shown in Figure 1c. Fluorescence photons were collected by the oil immersion objective (NA 1.4, 60×, Nikon DIC H), focused through a 150 um confocal pinhole. Single-photon events were detected by an avalanche photodiode (APD, EG&G SPCM-AQ-141) and arrival times (12.5 ns resolution) were recorded by the FPGA.

On-line control: The beam scan and feedback are generated and processed, respectively, on the FPGA board using a custom Labview program. This routine applies feedback voltage to the $x$- and $y$-electrode pairs based on the particle position that is calculated after each complete beam scan, based on the position of maximum absolute scattering contrast within the processed frame.

In order to reconstruct the scattering image prior to applying a trapping feedback force, at each scan position the signal measured by the photodiode is first digitized by taking the average of 8 rapid ADC reads during one 18.75 μs beam dwell time, Figure 2a. Once the signal at each scan position is recorded, a single interferometric scattering frame is reconstructed as described in the text, where each pixel is assigned the recorded raw signal value from the corresponding scan location (Figure 2b). The feedback voltage is calculated and applied, closing the loop.

## Note S3: Analysis notes

Calculation of interferometric scattering contrast: Each raw ISABEL frame is reconstructed using the known scan timing and initial start position of the excitation beam. The values at each pixel represent the ADC counts, or equivalently, the measured intensity at the photodiode. During each trapping run, the user manually takes a new background measurement periodically (while nothing is in the trap), and the times of these background measurements are recorded. Therefore, during analysis, the background taken at these times is subtracted from all following frames, until the next background replaces it.

<u>Frame reconstruction – fluorescence:</u> For brightness determination, photon time-tags from the APD record were binned into 10 ms intervals, associated with the appropriate positions on the grid depending upon their arrival time relative to the motion of the scanning beam.

<u>Correction for power fluctuations in laser illumination:</u> Small fluctuations in laser power (< 1%) can affect the apparent interferometric scattering contrast. Therefore, prior to the background-subtraction step as described above, fluctuations in the measured intensity due to changes in the overall illumination intensity must be removed. This is accomplished by correcting the signal in each frame by the average value of the outer 16 pixels (Figure S1b). The measured values of these pixels are expected to reflect the background level whether or not a bead is trapped, and therefore any change in their measured intensity is taken to reflect an overall change in illumination intensity.

**Note S4: Noise**
Power spectral density plots for the ISABEL trap signal both in the absence (background) and presence (trapping) of a trapped object are shown in SI Figure S2. This figure also provides a direct comparison of the analysis protocol that utilizes only the value at the center pixel (Figure S2a) as compared to taking the maximum value over the local trap region (Figure S2b). The noise in both signals is highly similar; we take this as a good indication that either analysis method would be appropriate for use in calculating the trapping contrast.

**Note S5: Calculating the trapping contrast for various particle sizes (Figure 4)**
The contrasts for scaling versus particle size are calculated based on an event-by-event contrast level approach. This avoids biasing the analysis towards longer trapping events, as would be the case for calculations based on the histogram of contrast levels for all time. The events were identified based on the absolute fractional contrast time-trace, averaged for every 20 frames with a level-finding algorithm. The level-finding algorithm used a maximum likelihood change-point algorithm using a Gaussian noise model.[3] Consecutive levels within 10% of each other were combined. Levels were further filtered into trapping events based on level durations, feedback states and detected positions. Three requirements were set: 1) the level must last longer than 200 ms; 2) the level must have the feedback enabled with the correct polarity for at least the duration of the level; 3) the root mean squared deviation of the position of the particle with respect to the trap center over the level must be less than 1.5 grid points. The histograms were fit to a cubed-Gaussian model, with the probability density

$$p(x) = A \, x^{-2/3} \exp\left(-\frac{\left(x^{1/3} - x_0^{1/3}\right)^2}{2\sigma^2}\right)$$

based on the contrast scaling as the cube of a Gaussian-distributed size, see Fig S5. The mode and half-width-half-max of the fitted function were used as measures of the mean and the heterogeneity of the contrast.

The contrast was plotted against measured TEM diameters for gold nanoparticles nominally of 20, 30, 40, 50, and 60 nm. The error bars represent the sample heterogeneity. The data were fit in log-log space with a linear regression, representing power laws, with least-squares fitting. The slopes of the best fits were fixed to power law exponents of 3 (cubic) and 6 (hexic).

# Supplementary Figures

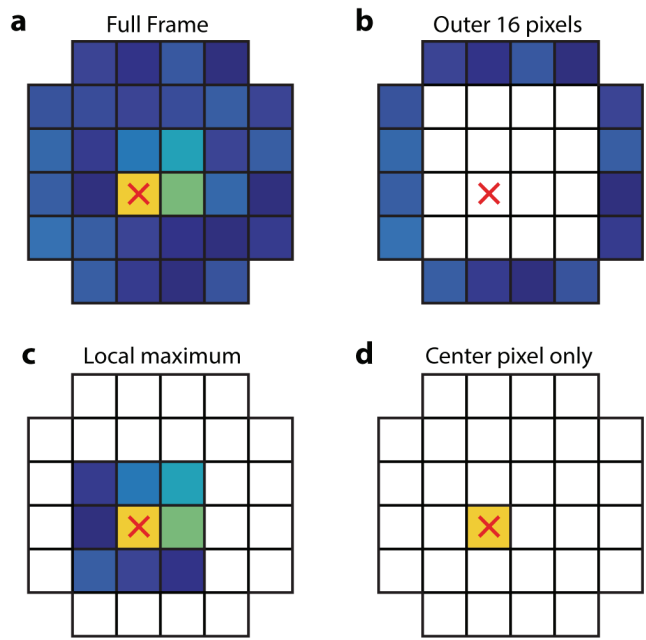

**SI Figure S1: Various pixel combinations used for analysis.** a) Full frame, with pixel pitch 0.5 μm and 32 pixel grid. b) Outer 16 pixels, used for correction of signal fluctuations due to excitation laser power. c) Local maximum of 9 pixels at and around the trap center is the value and location selected in each frame as most likely to represent a particle in that area. d) The absolute fractional scattering contrast at a single center pixel.

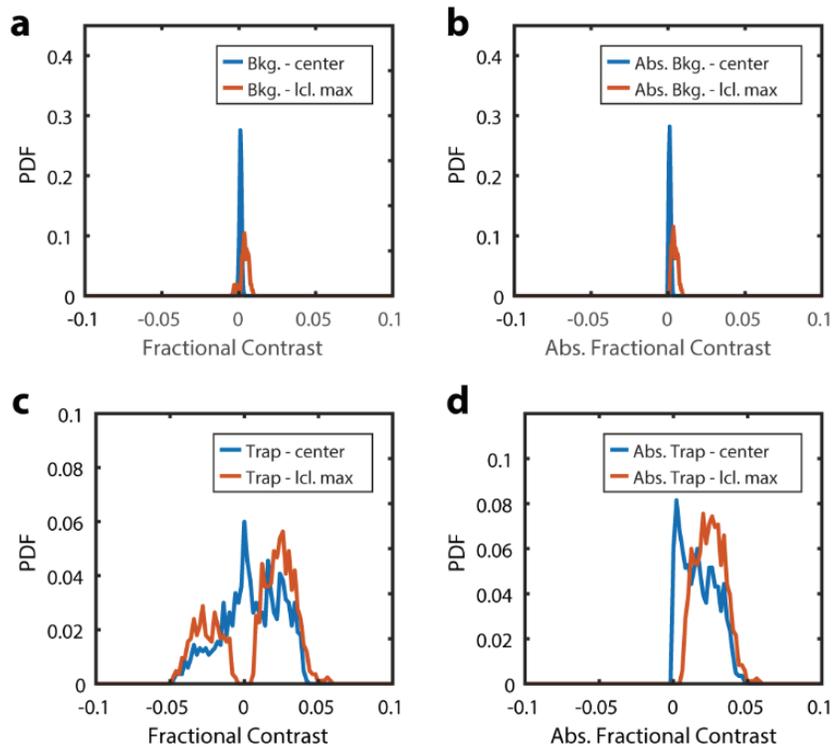

**SI Figure S2: ISABEL signal and background.** Comparison of the fractional contrast when calculated as the local max of the 9-pixel region (orange) or as the single central pixel (blue). Histograms are shown for background (a and b) vs. for trapped particles (c and d), as well as for a bipolar value of contrast (left panels) vs. absolute value (right).

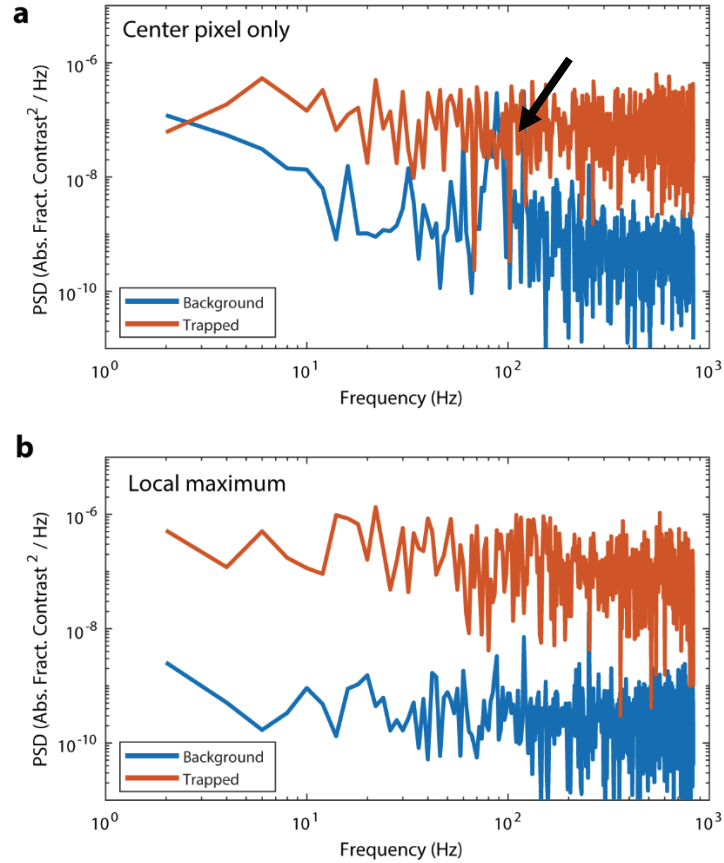

**SI Figure S3: Frame-by-frame noise in ISABEL signal and background.** a) Power spectral density of the absolute fractional contrast interferometric scattering signal taken only at the center pixel of the trap as shown in Figure S1d, compared to the power spectral density of the background signal at that pixel. b) Power spectral density of the absolute fractional contrast interferometric scattering signal taken as the maximum within the 9-pixel region shown in Figure S1c, compared to the power spectral density of the background signal from that region. Note that the peak in the background (blue trace, black arrow) taken at the center pixel only in a), along with low-frequency noise, is strongly suppressed when the local maximum is taken as in b).

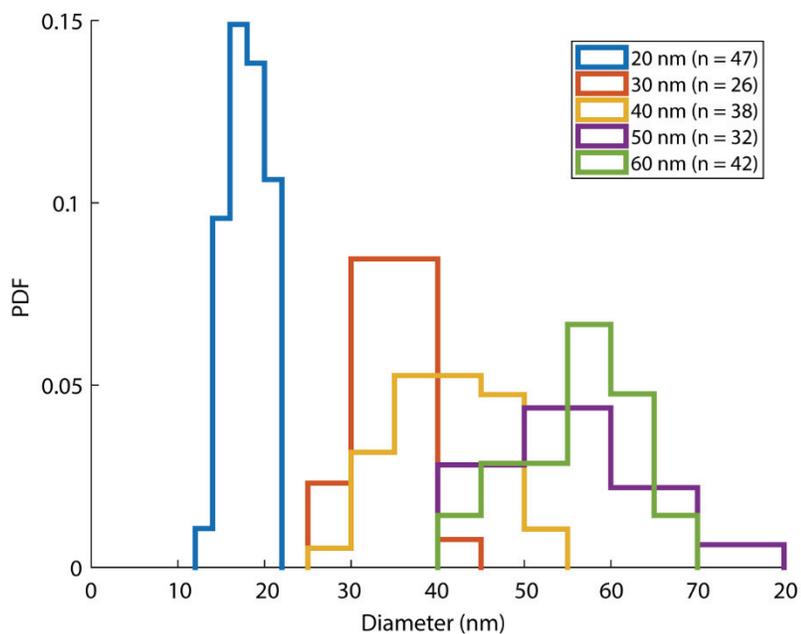

**SI Figure S4: Distribution of gold nanoparticle sizes.** Histograms of the diameters calculated from the measured area in TEM images as described in Note S1 for gold nanoparticles with nominal sizes of 20, 30, 40, 50, and 60 nm, plotted as a probability density function. The number of beads measured for each condition is in brackets.

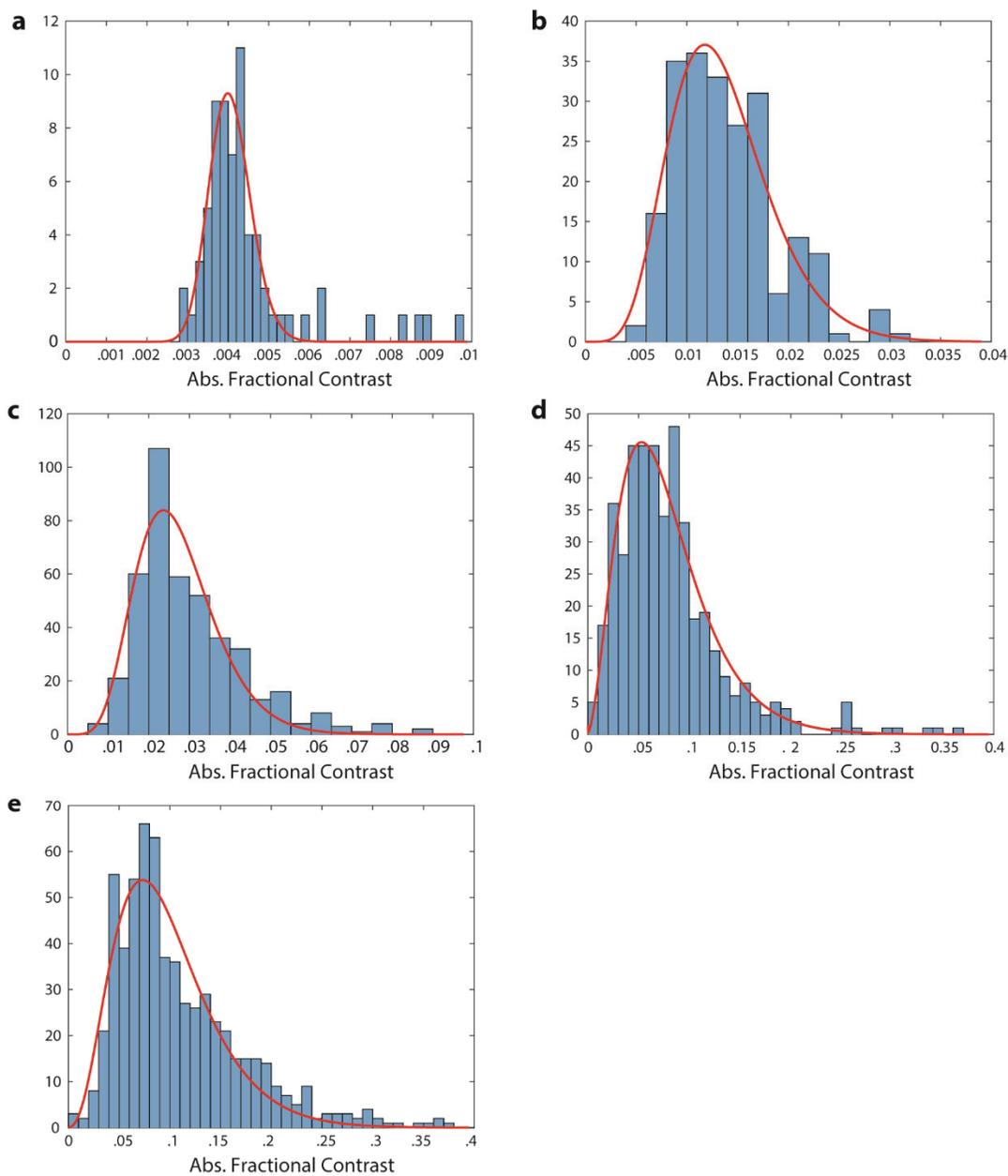

**SI Figure S5: Distribution of measured contrasts for gold nanoparticles.** (a-e) Histograms of the average absolute fractional contrasts for ISABEL trapping events for gold nanoparticles with nominal sizes of 20, 30, 40, 50, and 60 nm, respectively. Cubed Gaussian fits are overlaid.

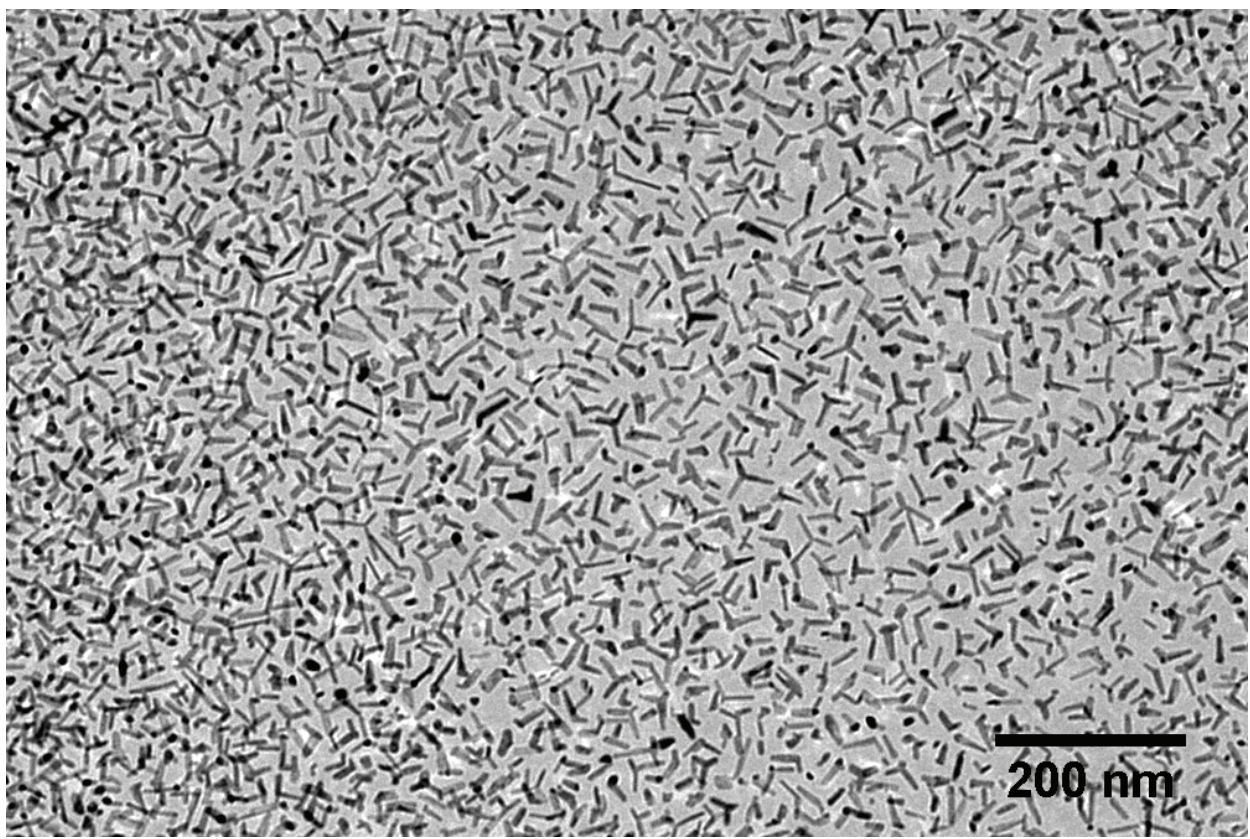

**SI Figure S6: Cryogenic TEM images of CdSe/CdS semiconductor nanoparticles.** Particles were imaged in vitreous ice by cryogenic electron microscopy following plunge freezing on Quantifoil Holey Carbon R 2/2 grids.

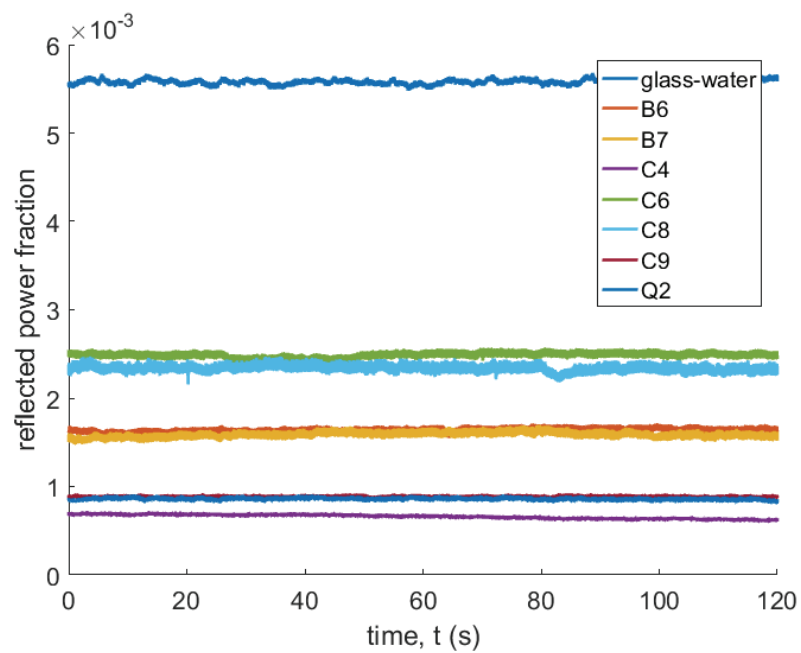

**SI Figure S7: Individual quartz microfluidic cells exhibit different reflectivity.** Each microfluidic cell exhibits a unique reflectance in the ISABEL trap, likely due to slight differences in fabrication. The reflected power from each of seven different cells (B6, B7, C4, C6, C8, C9, Q2) was monitored for two minutes in water on the ISABEL trap. The reflected power from a glass-water interface alone is provided for comparison.

## Supplementary Videos

Only screenshots are shown here; full videos are available for download.

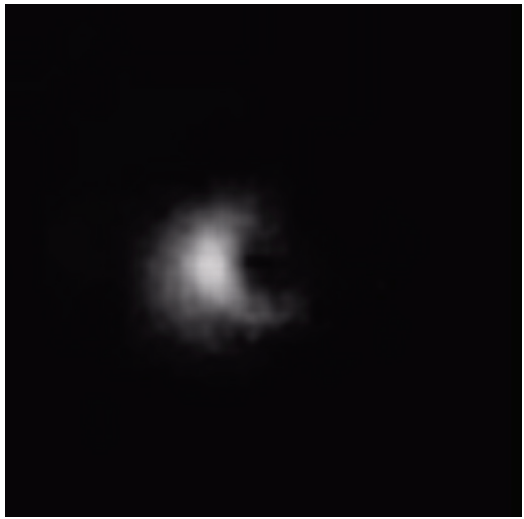

**Video S-1: Beam scan pattern (slow) for ISABEL Trap.** AODs are used to scan the focused excitation beam across the grid depicted in Figure 1. This video shows a ~ 10,000x slowed-down beam scan, taken with a conventional CMOS camera (TheImagingSource DFK 42BUC03). Two 0.2 μm polystyrene beads are visible as darker objects in the lower right.

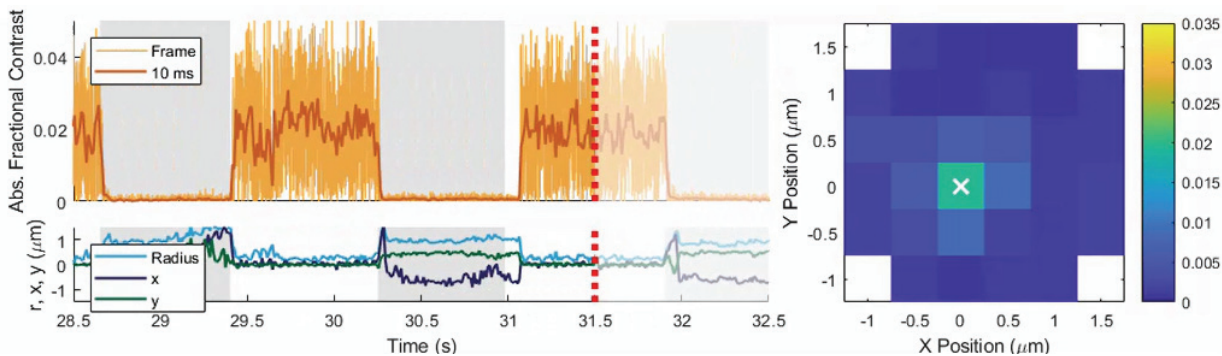

**Video S-2: ISABEL Trap with 40 nm gold beads (from Figure 3).** ***Top left:*** ISABEL absolute fractional contrast signal from the area around trap center, calculated as described in SI Methods above, and shown both frame-by-frame (light orange) and in 10 ms bins (dark orange). ***Bottom left:*** Algorithm-estimated object position, shown as radial distance from trap center (cyan), x position (navy), and y position (dark green). When the object is trapped, all three values are nearly zero. In both the top and bottom left-side panels, gray background indicates periods when feedback was OFF, white background indicates feedback ON. Red dotted lines indicate the point in time synchronized to the two right-side panels. ***Right:*** Reconstructed 32-pixel movie of ISABEL trapping, using the absolute fractional contrast at each beam scan position (colorbar). The trap center is marked at (0,0) as a white "x".

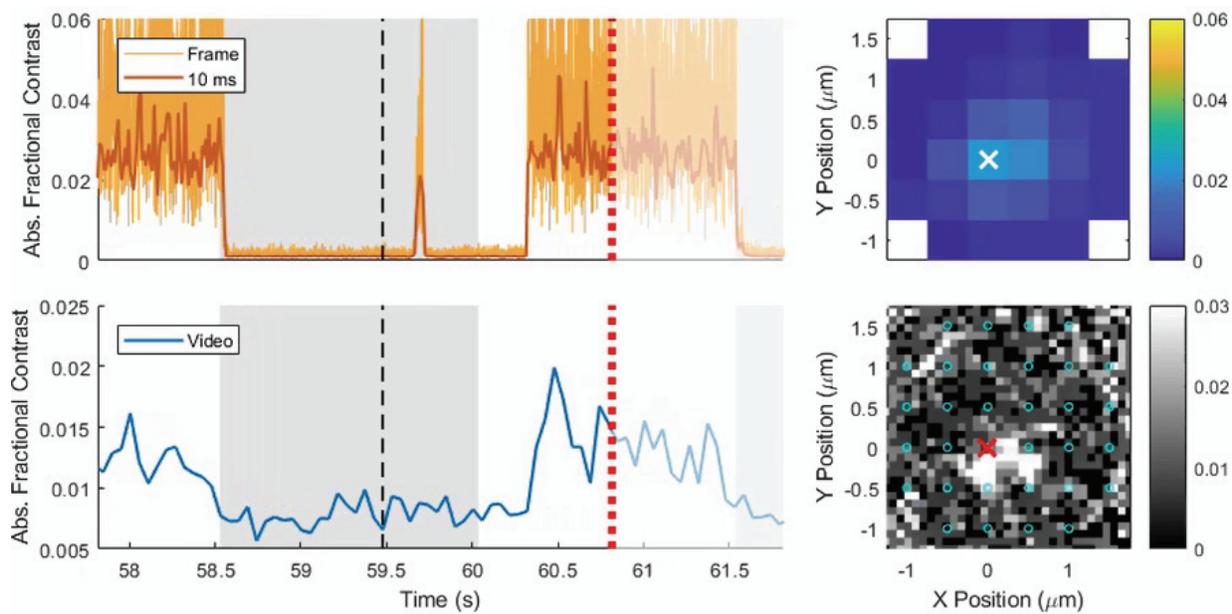

**Video S-3: Simultaneous flat-fielded video recording of 40 nm gold beads in the ISABEL Trap.** ***Top left:*** ISABEL absolute fractional contrast signal from the area around trap center, calculated as described in SI Methods above, and shown both frame-by-frame (light orange) and in 10 ms bins (dark orange). ***Bottom left:*** Video-based absolute contrast signal, acquired and calculated as described in SI Methods above. In both the top and bottom left-side panels, gray background indicates periods when feedback was OFF, white background indicates feedback ON. Red dotted lines indicate the point in time synchronized to the two right-side panels. ***Top right:*** Reconstructed 32-pixel movie of ISABEL trapping, using the absolute fractional contrast at each beam scan position (colorbar). The trap center is marked at (0,0) as a white "x". ***Bottom right:*** Flat-fielded video of the trapping region during acquisition, processed as described in SI Methods above. Approximate beam scan positions are indicated with small cyan circles. The trap center is marked at (0,0) as a red "x".

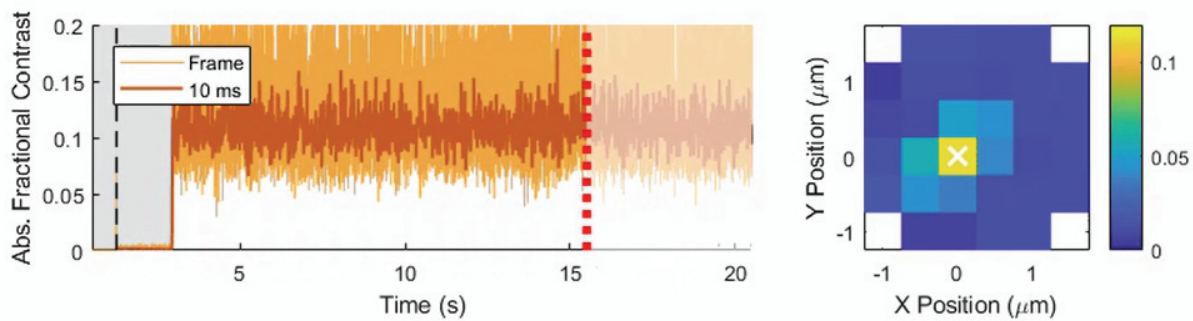

**Video S-4: Minutes-long trapping of a single 50 nm gold bead.** Single nanoscale gold beads can be trapped for minutes at a time. This example trace is sped up to 10x live speed, and shows a 50 nm particle trapped for more than two minutes before feedback is turned off to release the particle. **_Left:_** ISABEL absolute fractional contrast signal from the area around trap center, calculated as described in SI Methods above, and shown both frame-by-frame (light orange) and in 10 ms bins (dark orange). Gray background indicates periods when feedback was OFF, white background indicates feedback ON. Red dotted lines indicate the point in time synchronized to the right-side panel. **_Right:_** Reconstructed 32-pixel movie of ISABEL trapping, using the absolute fractional contrast at each beam scan position (colorbar). The trap center is marked at (0,0) as a white "x".

# Supplementary References